\documentclass[10pt,letterpaper,twocolumn]{article} %% two column, final layout

\usepackage{OL}
\usepackage{hyperref}
\usepackage{amsmath}
\DeclareTextSymbol{\degre}{T1}{6}
\DeclareTextSymbol{\degre}{OT1}{23}
\newcommand{\be}{\begin{eqnarray}}
\newcommand{\ee}{\end{eqnarray}}

\newcommand{\ket}[1]{\left| #1 \right\rangle}

\begin{document}

\twocolumn[ %% activate for two-column option

\title{Femtosecond spectral electric field reconstruction using coherent transients}

\author{Antoine Monmayrant, B\'eatrice Chatel, Bertrand Girard,}

\address{Laboratoire Collisions, Agr\'egats, R\'eactivit\'e (UMR 5589, CNRS - Universit\'e Paul Sabatier Toulouse 3), IRSAMC, France }

% Do not use \email or \homepage here. E-mail and URL can be given just before references.

\begin{abstract}We have implemented a new approach for
measuring the time-dependent intensity and phase of ultrashort
optical pulses. It is based on the interaction between shaped
pulses and atoms, leading to coherent transients.\end{abstract}

\ocis{(320.5540) Pulse shaping; (320.7100) Ultrafast measurements}

 ] %% activate for two-column option

\noindent

%\subsection{Introduction}

Complete pulse measurement is a central issue to most ultrashort
experiments. This is the case both for measuring the shortest
``clean'' pulses as well as complex shapes produced by pulse
shapers. The usual methods can be divided into self-referenced
techniques (SPIDER, FROG or derivatives)
\cite{dorrerchar01,iaconis98,Trebino1997} and techniques comparing
the unknown pulse with a well-known one. In the first method,
non-stationary filter or a non-linear interaction are required. In
the second case, the unknown field is compared to a known
reference field. A spectacular demonstration was achieved with
attosecond pulses used for direct light wave measurement
\cite{Krausz04_pulse_meas}. The method has a linear sensitivity
with respect to the unknown pulse, but previous knowledge of the
reference pulse can be a severe constraint, particularly in
examples such as spectral interferometry where the reference pulse
has similar characteristics (same spectral range, and similar or
broader spectral width) as the unknown pulse.

We present in this letter a new approach to electric field
reconstruction based on an atomic response. In this context, we
regard the atomic system as a high spectral resolution probe
(compared to spectrometers used in usual methods) of known response
to a shaped light field. Contrary to most of the well-known
cross-correlation in gases which require no resonant relay-state and
leads to intensimetric cross-correlation (just as in frequency
mixing processes)\cite{schins-JOSAB-96,Papadogiannis_OL02}, this
method requires resonant bound states and leads to the product of
the two electric fields. It is based on a sequence of Coherent
Transients (CT) measurements\cite{zamith01}. These CT are
oscillations in the excited state population resulting from the
interaction between a two-level system and a weak chirped pulse.
This scheme takes advantage of the high sensitivity of CT to the
pulse shape, in particular, to spectral phase effects
\cite{Degert02CTshaped,RbShapingAPB04}.
 In this letter, this new method is first explained
 then experimental results are presented. Its feasibility is illustrated
  by measuring glass material dispersion.
The limits of the method and its implementation are discussed.

%This method is particularly well adapted for weak chirped pulses.
%Indeed, an accurate characterization of the chirp profile is
%essential for many experimental applications such as pump-probe
%spectroscopy and coherent control. This method is based on
%pump-probe scheme and the experimental demonstration is made in
%the near infrared but could be extended to other spectral ranges
%by simply changing the two-level system.

%\subsection{Principle}
%Before presenting the characterization retrieval method, we recall
%the principle of the CT.
%let us consider a three
%level system $|g>, |e>, |f>$ excited by the common pump-probe
%scheme. In this one, one of the two pulses will be the unknown
%pulse $E(t)$, the other one being the reference pulse
%$E_{ref}(t)$. Let us consider first that the unknown pulse is the
%pump which creates a dynamic in the $|e>$, probed by $E_{ref}(t)$.
%The system is thus transported in the $|f>$ level. The
%experimental signal (fluorescence, ions..)observed function of the
%delay between pump and probe possesses all the dynamic information
%of the |e>-state. For a sake of simplicity, let us
%consider that the probe is short enough to be neglected, then
%focus our understanding on the method used to retrieve the unknown
%pulse from the dynamic of the population of the $|e>$state. This
%reduces the problem to
 %Let us consider
 The CT result from the interaction of a two-level system
($\ket{g}$ and $\ket{e}$) with a chirped pulse $E(t)$ of carrier
angular frequency $\omega_{0}$ close to resonance
($\omega_{0}\simeq\omega_{eg}$). The transient excited state
population is probed towards the $\ket{f}$ level in real time by a
second ultrashort pulse $E_{ref}(t)$ which is Fourier transform
limited and short compared to the characteristic features of $E(t)$.
Its frequency is close to resonance ($\omega_{fe}$). The
fluorescence arising from the $\ket{f}$ state is then recorded as a
function of the pump-probe delay $\tau$. The probe pulse provides
access to the temporal evolution of the population in $\ket{e}$,
produced by the pump beam. The result of the interaction is
described by first order perturbation theory, and the fluorescence
is proportional to
\begin{eqnarray}
S(\tau)&=&|a_f(\tau)|^2\\\nonumber &\propto&
\left|\int_{-\infty}^{+\infty}
E_{ref}(t-\tau)\exp(i\omega_{fe}(t-\tau)) a_e(t)dt \right|^2
\end{eqnarray}
%\mu_{eg}\mu_{fe}/4\hbar^2
 with
\begin{equation}\label{a_e}
a_e(t)= \int_{-\infty}^t E(t')\exp(i\omega_{eg}t')dt'
\end{equation}
In the case of a simply chirped pulse $E(t)$, a quadratic phase
appears in the integral giving $a_e(t)$ (Eq. \ref{a_e}), leading to
oscillations of the probability $|a_f(\tau)|^2$ as already
demonstrated \cite{zamith01,Degert02CTshaped}. These strong
oscillations result from interferences between the population
amplitude excited at resonance and after resonance. They are
extremely sensitive to tiny phase modifications
\cite{RbShapingAPB04}. However, although sensitive to phase effects
these CT give access to the excited state probability
$|a_e(\tau)|^2$ whereas the probability amplitude is necessary to
achieve a complete measurement of the electric field. Moreover, the
oscillations are only produced by the second part of the pulse
(after resonance)\cite{zamith01}.

To overcome these limitations, a new excitation scheme with two pump
pulses is used. Two measurements are performed, each with a two
pulse sequence with a well defined phase relationship
$E_{shaped}(t)=E_1(t)+e^{i\theta}E_2(t)$ where $E_1(t)$ and $E_2(t)$
are two replica of the unknown pulse generated by splitting the same
initial pulse $E(t)$ and adding additional spectral phase. These can
be obtained either with a Michelson-type interferometer or with a
pulse shaper. The first pulse $E_1(t)$ creates an initial population
in the excited state. The second pulse $E_2(t)$ is strongly chirped
and sufficiently delayed by
%$\tau$
$T$ so that it does not overlap with the first pulse. This second
pulse creates a population in the excited state which interferes
with the initial population created by the first pulse. Thus,
oscillations due to CT occur on the whole duration of the second
pulse. The final state population during the second pulse can be
written as
\begin{eqnarray}
|a_{f,\theta}(\tau)|^2&=& \left| a_{f,1}(+\infty)+e^{i\theta}a_{f,2}
(\tau)\right|^2 \nonumber\\
&=& \left| a_{f,1}(+\infty)\right|^2+ \left|a_{f,2} (\tau)\right|^2\nonumber\\
& & + 2Re\left\{a^*_{f,1}(+\infty)a_{f,2}(\tau)e^{i\theta}\right\}
\label{a_f}
\end{eqnarray}
$\left| a_{f,1}(+\infty)\right|^2$ can be obtained from the plateau
reached between the two pulses. The last two contributions in Eq.
\ref{a_f} depend on $\tau$ and on the second pulse. For a second
pulse of smaller peak power, the crossed term is dominant so that
the response is mostly linear with respect to the second pulse. A
first measurement for $\theta=0$ gives $ Re
\left\{a^*_{f,1}(+\infty)a_{f,2}(\tau)\right\}$. A second
measurement for $\theta=\pi / 2$ brings the complementary part $Im
\left\{a^*_{f,1}(+\infty)a_{f,2}(\tau)\right\}$. For pulse
intensities of comparable magnitude, the system of nonlinear
equations resulting from both measurements can be solved to extract
$a_{f,2}(\tau)$ \cite{MonmayrantCT-spirograph-04}. By derivation one
obtains

%the interferometric term which leads by derivation and Fourier transform to
% This
%provides with the contribution of the second pulse $a_2(t)$.
% Combining the results from these two scans allows one to extract the contribution from the second pulse $a_{e,2}(t)$ even when intensities are of comparable magnitude, which requires solving a non linear equation.
%Then, the time-dependent phase and amplitude of the pulse can be
%reconstructed by derivation. The experiment leads thus to the
%temporal measurement of $E_2(t) e^{-i\omega_{eg}t}$. One can
%retrieve $\widetilde{E_2}(\omega+\omega_{eg})$
% by fast Fourier transform.

%A probe pulse tuned on the $\ket{e}\rightarrow\ket{f}$ transition
%is used to measure the population $|a_e(t)|^2$ in real time.  Ideally, the probe
%pulse should be infinitely short. In practice, as in any
%referenced measurement its finite duration limits the time
%solution and therefore the frequency extent. More precisely,the
%fluorescence signal is proportional to
%\begin{equation}
%S(\tau)=|\mu_{eg}\mu_{ef}\int_{-\infty}^{+\infty}
%E_{probe}(t-\tau)\exp(-i\omega_{fe}(t-\tau))a_e(t)dt|
%\end{equation}$ $ and it can
%be shown that the quantity measured is the product of the two electric fields $
\begin{equation}\label{da_f}
\frac{{da_{f,2} }}{{d\tau }} = \int_{-\infty}^{+\infty} {
\widetilde{E}_2(\omega_{eg}+\Omega)\widetilde{E}_{ref}(\omega_{fe}-\Omega)d\Omega}
\end{equation}
from which $\widetilde{E}_2(\omega)$ can be deduced provided that
the reference pulse is known and short enough. Finally, knowing the
extra phase added to generate the pulse sequence,
$\widetilde{E}(\omega)$ and thus $E(t)$ is obtained. Note that
unlike conventional interferometric methods, the reference-field
spectrum does not need to overlap that of the unknown field.
\begin{figure}[htb]
\centerline{\includegraphics[width=9cm]{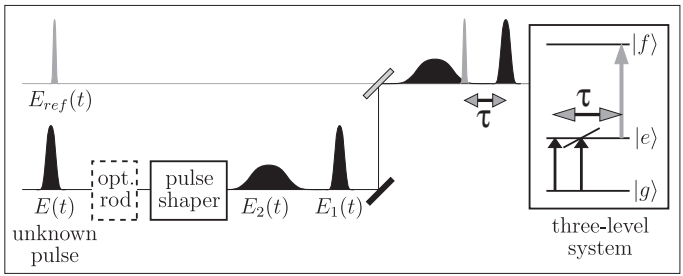}}
\caption{Set-up Principle:  The unknown pulse is sent into a pulse
shaper programmed to generate a sequence of two pulses. One FL, the
second delayed and chirped. An extra phase can be added. An optional
glass rod is added in the front of the set-up. The reference is a
pulse shorter than the unknown pulse. Inset: Excitation
scheme.}\label{setup}
\end{figure}
%\subsection{Experiments and results}

To illustrate the method, an experiment has been performed in an
atomic Rb vapor (see Fig. \ref{setup}). The Rb (5s - 5p $(P_{1/2})$)
transition (at 795 nm) is resonantly excited with the pump pulse
sequence. The transient excited state population is probed ``in real
time'' on the (5p - (8s, 6d)) transitions with an ultrashort pulse
produced by a home-made NOPA (607 nm, 25 fs). The ``unknown'' pulse
$E(t)$ has initially a duration of 130 fs and can be affected by
dispersive materials to demonstrate the measurement capabilities. A
phase and amplitude 640 pixels LCD-SLM pulse shaper
\cite{pulseshaperRSI04} is used to generate the pump pulse sequence
by applying a complex transmission in the spectral domain:
\begin{equation}
H_\theta(\omega)=\frac{\{1+\exp[i(\theta+\phi'(\omega-\omega_0)+\phi''\frac{(\omega-\omega_0)^2}{2}]\}}{2}
%H_\theta(\omega)=\frac{\{1+\exp[i(\theta+\phi'(\omega-\omega_0)+\phi''(\omega-\omega_0)^2/2]\}}{2}
\end{equation}
where $\omega_0$ is the carrier frequency of the pump pulse. The
first pulse in the pump sequence is identical to $E(t)$. The second
one is strongly chirped with $\phi''=-2.10^5 \;{\rm fs}^2$ in order
to produce CT, and delayed by $\phi'=6 \;{\rm ps}$. An extra phase
factor $\theta$ can be added.

\begin{figure}[htb]
\centerline{\includegraphics[width=9cm]{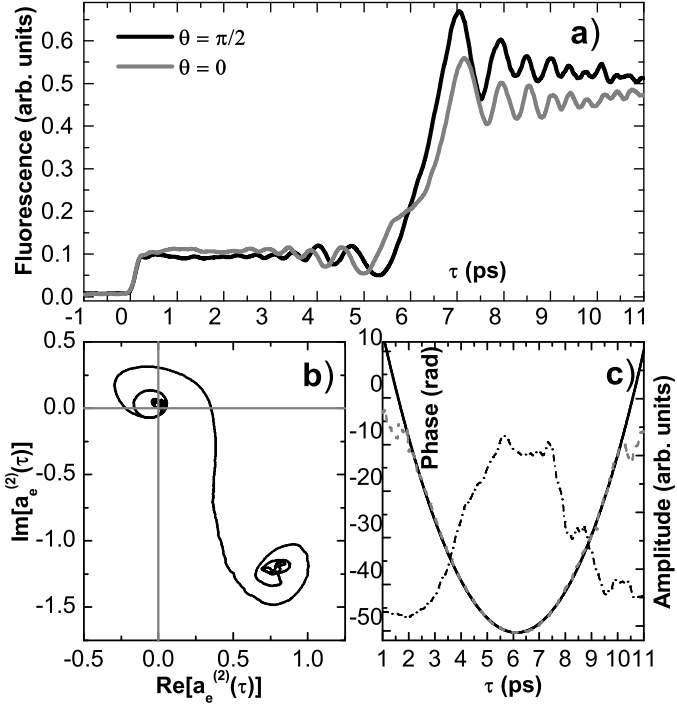}} \caption{a)
Experimental Coherent Transients resulting from the excitation of
the atom by a FT limited pulse (at time $\tau=0$) followed by a
chirped pulse (centered at $\tau=6$ps), for two different relative
phases $\theta =0$ and $\pi/2$ between the two pulses. b)
Probability amplitude $a_{f,2}(\tau)$ reconstructed from the two
measurements presented in a) and displayed in the complex plane. The
Cornu spiral appears clearly. c) Reconstructed phase (dots),
theoretical phase (dashed) and amplitude (dash-dotted) of
$E_2(\tau)$.}\label{FTresult}
\end{figure}

In a first experiment, $E(t)$ is close to Fourier limited. Two
recordings are performed for $\theta=\theta_0$ and
$\theta=\theta_0+\pi/2$ (with $\theta_0 \simeq -0.2\pi$) as shown in
Fig. \ref{FTresult}a. Combining these two measurements allows one to
determine in-phase and in-quadrature contributions from $E_2(t)$, so
that $a_{f,2}(t)$ can be retrieved. The main difference with
previous experiments \cite{zamith01,Degert02CTshaped} is the
preparation of a coherent superposition of $\ket{e}$ and $\ket{g}$
by the first pulse. Then the second -strongly chirped- pulse
produces large oscillations during its whole duration. These
oscillations can be seen as beats between the atomic dipole (which
behaves as a local oscillator) and the electric field from the
second pulse, as in heterodyne detection. By combining the two
measurements, it is therefore possible to retrieve fully the
temporal evolution of the excited state probability amplitude due to
the second pulse. Figure \ref{FTresult}b displays the reconstructed
excited state probability amplitude in the complex plane. The
expected Cornu spiral \cite{zamith01} is observed. $E_2(t)$ obtained
by simple derivation of $a_{e,2}(t)$, is displayed on Fig.
\ref{FTresult}c. The temporal amplitude and phase are represented.
As a comparison, the exact theoretical temporal phase applied by the
pulse shaper is shown (dashed line) without any other adjustment
than the offset. The agreement is excellent. The quadratic phase
added by the pulse shaper is perfectly retrieved.

\begin{figure}[htb]
\centerline{\includegraphics[width=9cm]{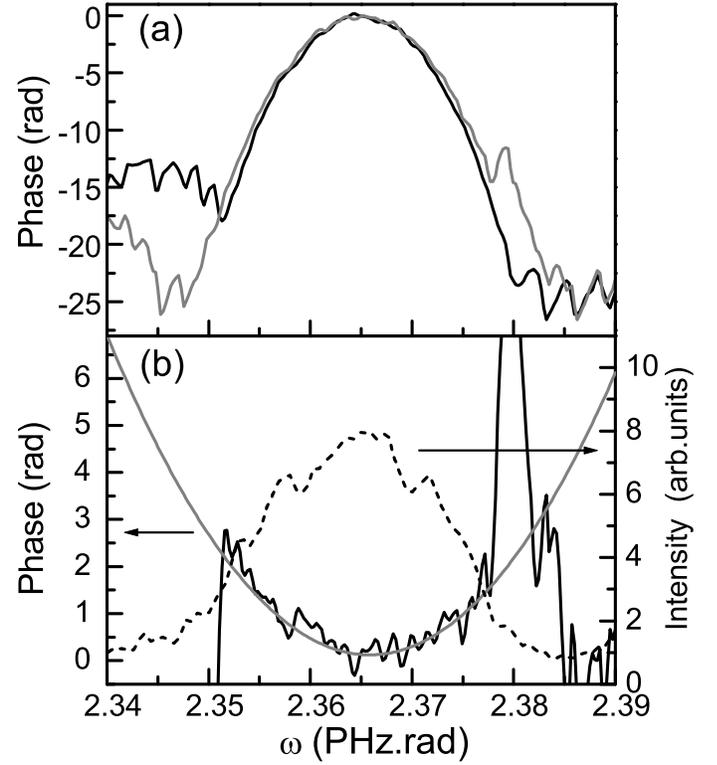}} \caption{Electric
field reconstruction. a) Spectral Phase retrieved obtained with
(gray) and without (black) rod. b) Dashed line: Spectral intensity;
Solid lines: Spectral phase due to the rod, (black): Experiment
(difference between data shown in a)), (gray): Calculated phase from
rod properties.}\label{EFresult}
%the amplitude of the electric field
%(squares) with a solid line gaussian fit. b) Spectral phase of
%$E^{(2)}$ (squares). The solid line is the theoretical phase
%taking into account the quadratic value introduced by the pulse
%shaper and the glass rod phase. c) Glass rod phase retrieved by
%substraction between the phase b) and the reference phase
%monitored without the rod (squares). Phase calculated (solid
%line).}
\end{figure}

 In a second set of experiments, the dispersion of a
SF58 glass rod ($\phi''=20 492 \, {\rm fs}^2$) inserted in the pump
beam is measured. CT are monitored with and without the rod in the
pump beam. This dispersion is sufficiently small so that $E_1(t)$
and $E_2(t)$ do not overlap. Experimental results in the spectral
domain are presented on Fig. \ref{EFresult}. Figure \ref{EFresult}a
shows  the spectral phase of $\widetilde{E}_2(\omega)$ with and
without the dispersive rod. Their difference is plotted on Fig.
\ref{EFresult}b together with the value calculated from the rod
coefficients. The agreement is excellent on the spectral domain
where the intensity (dashed line) is significant.
%A polynomial fit of
%the experimental data gives a 10\% mismatch on the quadratic
%phase.

 We have thus
demonstrated here the capabilities of this method on a simple
example. Its main advantage is that the unknown and reference pulses
need not have any spectral overlap, thus the method may be easily
extended to wavelength regions for which a local oscillator pulse is
not usually available. This method provides interferometric-like
terms
%$\widetilde{E}(\omega+\omega_{eg})\widetilde{E}_{ref}(\omega-\omega_{ef})$
but the two spectral shears remove constraints of usual
interferometric technics : same spectral domain, interferometric
control of delay, temporal step small enough to resolve fringes.
Here the heterodyne beating with the local oscillator provided by
the atom shifts the interferometric oscillations around zero
frequency which makes them much easier to measure. The requirement
of an atomic transition can be seen as a constraint as compared to
%makes this method less attractive than
"all optical" methods for pulse measurements in the visible or near
IR. However, some complex pulse shapes (with holes or phase
discontinuities) are difficult to measure with standard methods and
the present one can be more appropriate (numerical simulations are
very encouraging). Moreover this new method can present significant
advantages in other spectral range (UV, mid-IR). The condition is to
find a three-level system which allows a pump-probe scheme. The
pulse sequence could be implemented in a simpler set-up using a
modified Michelson interferometer instead of the pulse shaper.
%The
%additional quadratic phase allows to have better signal to noise
%ratio.
The spectral resolution here is not limited by that of the pulse
shaper but only by the temporal interval scanned and ultimately by
the linewidth of the atomic transition (typically $10^{-6}$), which
could be order of magnitude improvements with respect to other
methods based on a spectrometer. There is no intrinsic limitation
for the bandwidth or the central frequency of the unknown pulse
provided that a system with resonant transitions exist in this
range. The reference pulse bandwidth should be as large as this of
the unknown pulse. We sincerely acknowledge Christophe Dorrer and
Manuel Joffre for fruitful discussions and advices.

%The role of the chirp of the secund pulse is definitely not
%required for the feasibility of the experiment but allows to
%increase the sensitivity to any modification of the unknown phase.
%Consequently it is not a requirement but a wish.

%\nocite{*}
\bibliographystyle{osajnl}

\end{document}